# An electro-optically tunable arrayed waveguide grating fabricated on thin film lithium niobate

*Zhe Wang,[1] Zhiwei Fang,[1,]\* Yiran Zhu,[1,2] Jian Liu,[1] Lang Gao,[3] Jianping Yu,[1] Haisu Zhang,[1] Min Wang,[1] And Ya Cheng[1,2,3,4,5,6,]\**

Corresponding author: zwfang@phy.ecnu.edu.cn; ya.cheng@siom.ac.cn

[1] The Extreme Optoelectromechanics Laboratory (XXL), School of Physics and Electronic Science, East China Normal University, Shanghai 200241, China
[2] State Key Laboratory of Precision Spectroscopy, East China Normal University, Shanghai 200062, China
[3] State Key Laboratory of High Field Laser Physics and CAS Center for Excellence in Ultra-intense Laser Science, Shanghai Institute of Optics and Fine Mechanics (SIOM), Chinese Academy of Sciences (CAS), Shanghai 201800, China
[4] Collaborative Innovation Center of Extreme Optics, Shanxi University, Taiyuan 030006, China
[5] Hefei National Laboratory, Hefei 230088, China
[6] Joint Research Center of Light Manipulation Science and Photonic Integrated Chip of East China Normal University and Shandong Normal University, East China Normal University, Shanghai 200241, China

**Abstract:** We design and fabricate an 8-channel thin film lithium niobate (TFLN) arrayed-waveguide grating (AWG) and demonstrate the electro-optical tunability of the device. The monolithically integrated microelectrodes are designed for waveguides phase modulation and wavelength tunning. Experiments show that the fabricated electro-optically controlled TFLN AWG has a channel spacing of 200 GHz and a wavelength tuning efficiency of 10 pm/V.

## 1．Introduction

Optical wavelength-division multiplexing (WDM) is a powerful technique for fully exploring the bandwidth of an optical fiber for enabling high-capacity transmission in telecommunication systems. Specifically, the arrayed waveguide grating (AWG) is a key component for dense WDM system [1-3]. Generally, environmental factors such as temperature, vibration, and electromagnetic radiation can significantly affect the performance of AWGs. Therefore, precise wavelength control is necessary for these AWGs. So far, the control of AWGs is mainly achieved through thermo-optical effect due to the fact that the AWGs are mainly constructed on photonic integration platforms such as silicon on insulator (SOI), silica-based planar lightwave circuits (PLC), and silicon nitride [4-6]. In comparison with the thermal-optic effect, the electro-optic effect (i.e., Pockels effect), offers faster modulation speeds and lower power consumption. Nevertheless, the tunable AWG implemented using the electro-optic effect has not been achieved yet. Here, we demonstrate an electro-optically tunable AWG on thin film lithium niobate (TFLN). Notably, TFLN is an excellent photonics integration platform, which is well-known for its outstanding optical characteristics including a wide transmission window (0.35-5 μm), a relatively large refractive index (~2.2 @1.55 μm), a high second-order optical nonlinearity ($d_{33} \approx -27$ pm/V), and a low optical loss (~1 dB/m). Furthermore, the outstanding Pockels coefficient ($r_{33} \approx 31$ pm/V) of TFLN has made it widely adopted for electro-optic modulators [7-9]. Recently, AWGs on Z-cut and X-cut TFLN platforms have both been realized while the electro-optic tunability has not been achieved [10-14]. Herein, we design and fabricate 8-channel TFLN AWG and characterize the electro-optically tunability in the fabricated device. Notably, unlike the thermal-optic modulation which tunes the phase in the entire device, our electro-optic modulation can be conducted in each waveguide which offers a high flexibility and precision in terms of the spectral control and dispersion compensation.

## 2. Design and Fabrication

As shown in Fig.1(a), we design an electric-optic (EO) tunable TFLN AWG on the X-cut TFLN on insulator wafer. The TFLN AWG consists of an input waveguide, eight output waveguides, two free prorogation regions (FPRs), an array of twenty waveguides with fixed length difference, and an array of twenty pairs microelectrodes. The AWG satisfies the grating equation of $n_{eff} \times \Delta L = m\lambda$, where $n_{eff}$ is the effective refractive index of the fundamental transverse electrical (TE) mode, $\Delta L$ is the length difference between adjacent waveguides in waveguide array, $m$ is the diffraction order, and $\lambda$ is the center wavelength. On the X-cut TFLN, the optical axis lies within the wafer plane, so the planar anisotropy on the X-cut wafer makes it difficult for the design of bending waveguide. It is because the $n_{eff}$ of the TE mode in waveguide array varies with the angle $\theta$, which is the angle between the waveguide propagation direction and the crystalline Z-axis of TFLN on the wafer plane. That is why the TFLN AWGs so far mostly based on the Z-cut film, as the invariant refractive index regardless of the light propagation directions in the plane of TE polarization [10-13]. To better utilize the high electro-optic coefficient of the X-cut lithium niobate, J. Yi et al. designed and fabricated anisotropy-free dispersive components on a X-cut TFLN recently, which satisfies the equation of $n_{eff}(\theta) = n_{eff}(0°)\cos^2\theta + n_{eff}(90°)\sin^2\theta$ [14]. Therefore, the phase of the array waveguides on the X-cut TFLN can be accurately integrated along the bending waveguide. As a result, they fabricated the array waveguides along an angle of 45° on X-cut TFLN for the phase of the arrayed waveguides on both sides of the Z-axis cancel each other out. Based on this principle, we designed twenty array waveguides and a 1.6 nm (200 GHz) channel spacing with the length difference $\Delta L$ of 85.5 μm. The diameter of the Rowland circle is designed to be 610.6 μm, a linear adiabatic linear tapered waveguide is employed from the Rowland circle to the array waveguide to reduce coupling losses. To achieve electro-optic tunable AWG, a 1-cm-long electrodes array were designed along the straight section of the arrayed waveguides. As shown in inset of Fig.1(a), the electrodes are divided into signal electrodes and ground electrodes, the width of electrode is 10 μm, and the gap of electrodes is 5 μm. The signal electrodes in the twenty pairs of electrodes are distributed independently, while the ground electrodes are connected together on the chip.

Fig. 1(b) shows the cross-sectional schematic view of the TFLN waveguide and two electrodes. The low-loss optical waveguide features an TFLN ridge waveguide with a top width of 1 μm, a bottom width of 4.8 μm, and an etching depth of 210 nm. A 1-μm-thick $SiO_2$ cladding is on the top to protect the TFLN ridge waveguide. Fig.1(c) shows the numerically simulated overlap between the corresponding optical and electric fields. The gap between the electrodes on the two sides of the waveguide is 5 µm. This gap is designed to be as small as possible to achieve a low half-wave voltage while ensuring reasonably low propagation loss for the light traveling within the waveguide. It is crucial that the two electrodes are sufficiently far away from the optical mode in the waveguide. Fig.1(d) demonstrates the variation of the optical refractive index as a function of the applied electric field. The slope of the linear variation is $\frac{dn_{eff}}{dE} = 6.7244 \times 10^{-7}\ m/V$. Since the tuning efficiency is linearly distributed, we can precisely control the phase by changing the effective refractive index of the waveguide through voltage adjustments on both sides of the waveguide. The simulated transmission spectrum of the 1×8 TFLN AWG is also obtained as shown in Fig. 1(e), the channel spacing is 1.6 nm or 200 GHz around wavelength 1550 nm.

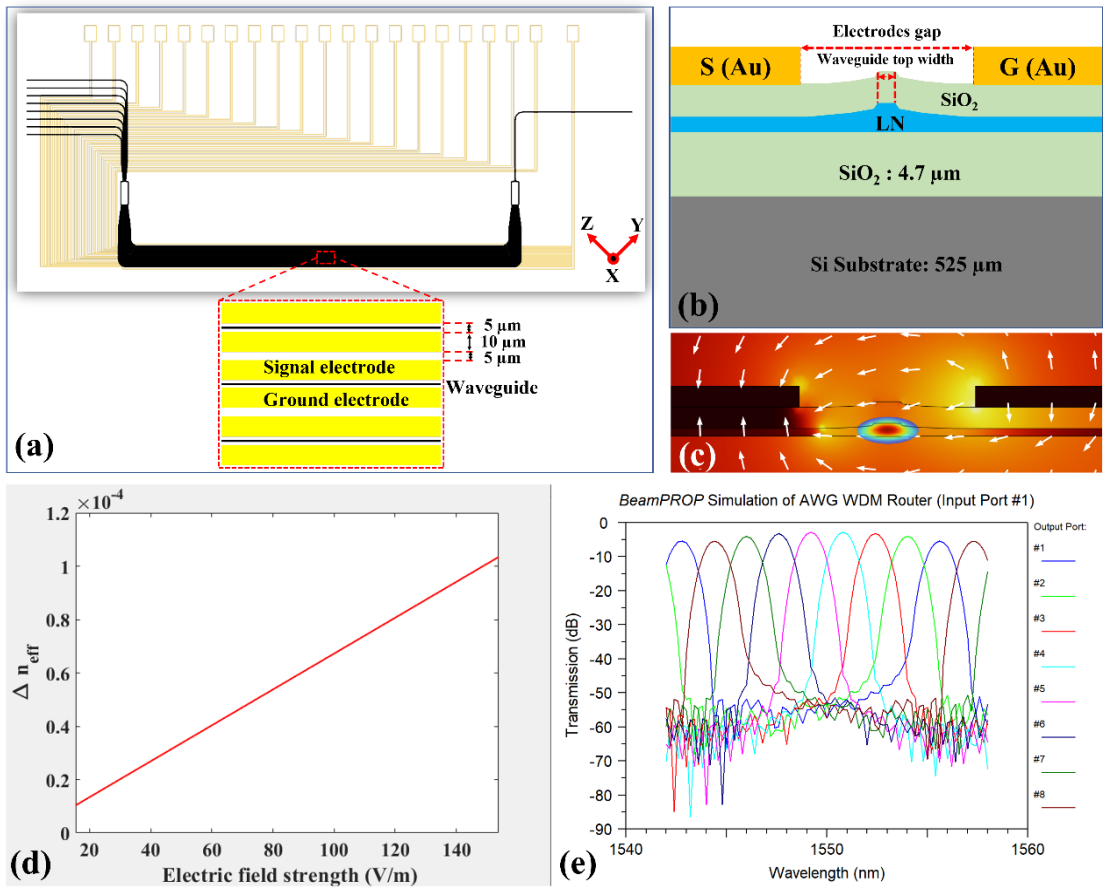

Figure 1. (a) Schematic diagram of EO tunable TFLN AWG, which consists of TFLN AWG and microelectrodes. (b) The cross-sectional schematic view of phase modulation

section which consists of the TFLN waveguide and two electrodes. (c) The cross-section view of the simulated optical TE mode profile in TFLN waveguide and electrical field (shown by white arrows). (d) The variation of the optical refractive index of TFLN waveguide as a function of the applied electric field. (e) The simulated transmission spectra of the different output ports of the 8 channels TFLN AWG around 1550 nm.

Fig.2 (a) shows the fully packaged EO tunable TFLN AWG. The TFLN AWG is fabricated on a 300-nm-thick X-cut TFLN on a $SiO_2$/Silicon (4.7 μm/525 μm) substrate (NANOLN) using photolithography assisted chemo-mechanical etching (PLACE) technique, the details about the TFLN waveguide device fabrication can be found in our previous works [15-17]. A 1-μm-thick cladding $SiO_2$ film was deposited on TFLN waveguide using plasma-enhanced chemical vapor deposition (PECVD). An array of metal electrodes (30 nm Cr/500 nm Au, deposited using a magnetron sputtering) on the cladding $SiO_2$ layer was fabricated along the straight waveguide. All electrodes are guided to the edge of the chip for easy connection with the printed circuit board (PCB) for wire bonding. The electrode pattern with a spacing of 300 μm is placed around the periphery of the TFLN AWG chip to match the pin spacing on a PCB board. After fixing the AWG sample on the PCB, gold wires are used to connect the electrodes between them. Fig. 2(c) is the zoom-in image of the region between FPR and array waveguides, the 50-μm-long adiabatic linear taper has a minimum width of 1 μm, a maximum width of 8 μm, a minimum gap of 0.5 μm, and a maximum gap of 8 μm, which achieves low coupling loss between the FPR and array waveguides. Fig. 2(c) is the Euler bend waveguides array which has a minimum radius of 200 μm. The loss of the Euler bend waveguide is several times smaller compared to the circular bend of the same effective radius at the telecom wavelength [18]. Fig. 2(d) shows the TFLN waveguide and the GS (Ground-Signal) electrodes on both sides, the gap of electrodes is about 5 μm, a smaller gap implies that a greater modulation efficiency can be achieved.

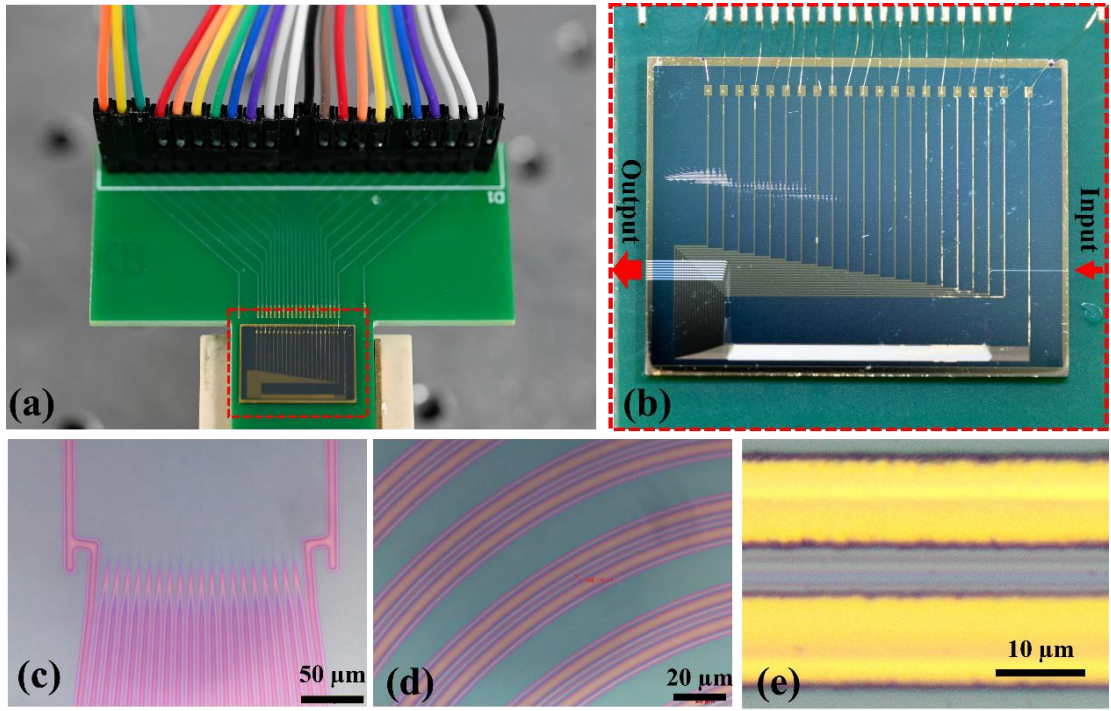

Figure 2. (a) Photograph of the fully packaged EO tunable TFLN AWG. (b)The micrographs of the fabricated 1 × 8 EO tunable TFLN AWG. (c)The magnified micrograph of the FPR. (d)The micrograph of the bend waveguides of array waveguides. (d)The micrograph of the straight waveguide and integrated microelectrodes.

## 3. Results and Discussions

Fig. 3(a) shows a schematic diagram of the experimental setup for the measurement of the EO tunable TFLN AWG. We fix the packaged EO tunable TFLN AWG chip on a on a fiber-chip coupling stage. A C-band continuously tunable laser (CTL 1550, TOPTICA Photonics Inc.) generates tunable laser signal around 1550 nm, which is injected into the input port of chip through a lensed fiber. The polarization of the signal laser is adjusted using an in-line fiber polarization controller (FPC561, Thorlabs Inc.). The optical signal at the output port of chip is connected to a photodetector (1611FC-AC, Newport) through another lensed fiber, and the photodetector receives the optical signal and connects to an oscilloscope (Tektronix MDO3104) through a cable. A multi-channel programmable direct current (DC) Power Supplies (IPMP250-1L, INTERLOCK) were used to provide DC voltage to the 20 pairs of electrodes on the chip through the PCB. The measured spectra of the TFLN AWG without applying DC voltage is shown in in Fig. 3(b). It can be observed that the center wavelength is 1550 nm and total-channel bandwidth is about 12.8 nm. The channel spacing is measured to be 1.6 nm which is well consistent with the simulation results.

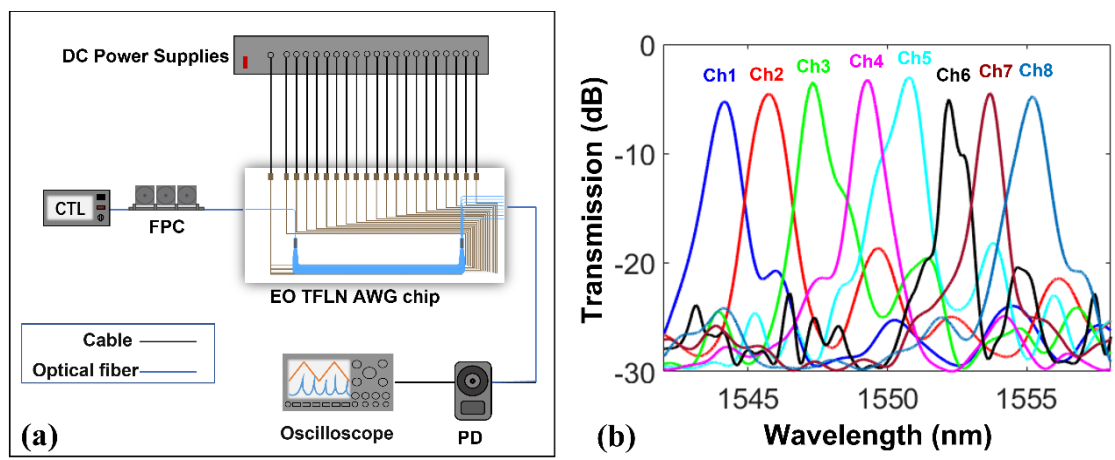

Figure 3. (a) Schematic diagram of the experimental setup. TCL: continuously tunable laser, FPC: fiber polarization controller, PD: photodetector. (b) The measured transmission spectra of the fabricated 8-channel TFLN AWG without applying DC voltage.

Then, we characterize the electro-optic tunable performance of the TFLN AWG. The transmission spectra of the 8 output channels are examined at various voltages applied to the electrodes using multi-channel programmable DC Power Supplies. Currently, all signal electrodes are set to the same DC voltage. To observe the tuning behavior of all channels with different voltages, we collected the spectra of all channels at 0V and 60V as shown in Fig. 4(a). We found that the entire spectrum undergoes a redshift as the voltage increases. As shown in Fig. 4(b), we collected the spectra of Channel-4 at voltage from 0V to 60V in steps of 10V. As depicted in Fig. 4(c), the wavelength shift exhibits nearly linear behavior with varying DC voltages, indicating excellent linearity in the electro-optic tuning process. It is observed that the center wavelength experiences a red shift of approximately 0.6 nm with increased 60 V, the slop reveals a tuning efficiency of 10 pm/V. It demonstrates the precisely control of the spectra through EO tunability.

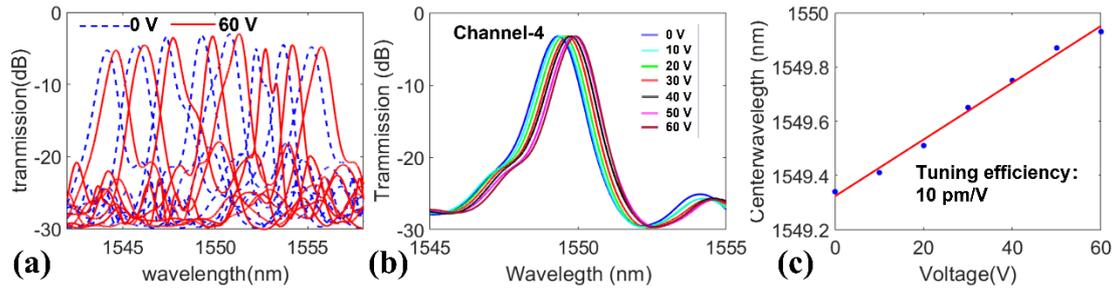

Figure 4. (a) The transmission spectra of 8 channels with applied 0 V and 60 V. (b) The transmission spectra of channel-4 from 0 V to 60 V, with intervals of 10 V. (c)The center wavelength shows a linear redshift as function of applied DC voltage.

## 4. Conclusions and outlook

We fabricated an 8-channel TFLN AWG with electro-optical tunability. The fabricated EO-tunable TFLN AWG has a channel spacing of 1.6 nm and the wavelength tuning efficiency of 10 pm/V. The device also features a low propagation loss and excellent tuning linearity. In the future our efforts will be devoted to achieving higher tuning efficiency and wider tuning range, making the device highly attractive for optical communications, frequency tuning and sensing, to name a few applications.

**Acknowledgements**
National Key R&D Program of China (2019YFA0705000), National Natural Science Foundation of China (12274133, 12004116, 12104159, 12192251, 11933005, 12134001, 61991444), Science and Technology Commission of Shanghai Municipality (21DZ1101500), Shanghai Sailing Program (21YF1410400). Innovation Program for Quantum Science and Technology (2021ZD0301403), the Fundamental Research Funds for the Central Universities (East China Normal University).

Received: ((will be filled in by the editorial staff))
Revised: ((will be filled in by the editorial staff))
Published online: ((will be filled in by the editorial staff))